\documentstyle[11pt,newpasp,twoside,epsf]{article}
\def\nodata{ ~$\cdots$~ }

\def\lae{\mathrel{<\kern-1.0em\lower0.9ex\hbox{$\sim$}}}
\def\gae{\mathrel{>\kern-1.0em\lower0.9ex\hbox{$\sim$}}}
\def\edcomment#1{\iffalse\marginpar{\raggedright\sl#1\/}\else\relax\fi}

\reversemarginpar

\begin{document}

\title{The Formation of Globular Cluster Systems}
\author{Patrick C\^ot\'e}
\affil{Rutgers University, 136 Frelinghuysen Road, Piscataway, NJ 
08854}

\begin{abstract}
I briefly review models for the formation of globular cluster (GC) systems,
and summarize the observational properties ($i.e.$, formation efficiencies,
metallicity distributions, kinematics and ages) of the GC systems of M87 and
M49: two thoroughly-studied elliptical galaxies. Many of the properties
of the metal-poor GCs in these and other galaxies appear to be consistent
with their formation in low-mass proto-galactic fragments, as proposed by 
several different formation models. A number of outstanding questions 
concerning the formation of the metal-rich GCs in these galaxies are highlighted.
\end{abstract}

\section{Introduction}

The past decade has witnessed a remarkable advance in our understanding of 
globular cluster (GC) systems. Optical and ultraviolet imaging from {\it HST} 
has been used to study GC systems belonging to nearly one hundred (mostly 
early type) galaxies. Spectra from 4-10m class telescopes have yielded
radial velocities for more than a thousand extragalactic GCs, along with 
metallicities for subset of the brighter objects. 
Infrared photometry for extragalactic GCs is appearing in the literature 
at an ever-increasing rate, promising to refine
our understanding of how these systems formed and evolved.

Indeed, theorists are now hard pressed to produce a single model
for the formation of GC systems that explains --- or, at least, is consistent 
with --- the enormous amount of observational material that has been accumulated.
This is a daunting task, since any viable formation model
must reproduce and explain the apparently constant number of GCs per unit baryonic 
mass, the striking complexity of GC metallicity distribution functions (MDFs), 
the observed spatial distributions of GCs about their host galaxy, and the 
emerging constraints on the ages and kinematics of the metal-poor (MP) and metal-rich 
(MR) GC subpopulations. What is ultimately required is a model that explains not 
just these observational properties, but also their variations (if any) with host 
galaxy environment and morphology. Although several important issues remain 
unresolved, I will argue below that there may be an emerging consensus on 
the origin of the MP GC subpopulations associated with bright galaxies. The
origin of the MR subpopulation remains very much open to debate.

\section{Formation Scenarios}

It is convenient to divide models for the formation of 
GC systems into three broad categories: pre-, proto- and 
post-galactic. While these divisions are useful for heuristic
purposes, they should not be taken too literally since the classification
is, in some cases, rather subjective. It is also worth noting that some models
that have historically been considered quite distinct ($e.g.$, Ashman \&
Zepf 1992 and C\^ot\'e $et~al.$ 1998) actually rely on
similar mechanisms  --- mergers and accretions --- to explain 
the formation of GC systems.
These models differ mainly the assumed nature of the
merger/accretion process: most notably in the 
assumed mass spectrum of the progenitors, the epoch of galaxy
assembly, and the amount of dissipation and star/GC formation (if any)
induced by the merger/accretion process. 

\subsubsection{\bf Pre-galactic:} In this picture, the formation of GCs predates 
that of their host galaxy. First suggested by Dicke \& Peebles (1968),
who noted that the Jeans mass at recombination was similar to the mean mass of 
Galactic GCs, this model was later revised by Peebles (1984) to explain GC 
formation in a cold dark matter (CDM) dominated universe. We now 
recognize the GC mass distribution to have a roughly double power-law form
with a break at $M \sim 10^5~M_{\odot}$, weakening the original Jeans 
mass argument.

Recently, Cen (2001) has proposed a variant of this model in which the MP
GCs surrounding galaxies form during reionization, when the external radiation 
field produces inward shocks in low-mass, low-spin dark matter halos. Such shocks, 
he argues, will compress the baryons and lead to the formation of GCs. While this 
model gives roughly the correct mass spectrum for GCs more massive than about 
$10^5~M_{\odot}$, the mechanism itself is quite unlike that observed 
in star- and cluster-forming molecular clouds in the local universe 
(and leaves the formation of the MR GCs unexplained). Moreover,
it seems inevitable that this model should predict many intergalactic GCs in
the local universe. Although the properties of some GCs associated with 
a few galaxies like M87 are consistent with an intergalactic
origin (see C\^ot\'e $et~al.$ 2001), there is still no case in which 
even a single GC can be said with certainty to be truly intergalactic 
in nature.

\subsubsection{\bf Proto-galactic:} 

\begin{figure}[t]
\plotfiddle{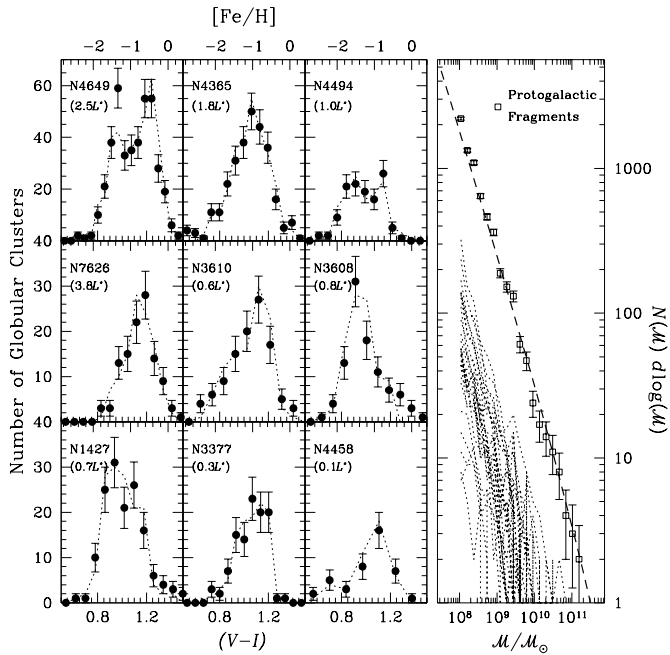}{3in}{0}{45}{45}{-150pt}{-80pt}
\caption{(Left Panels) Representative color/metallicity distributions for GCs
belonging to nine of the 28 galaxies in the study of Kundu \& Whitmore (2001).
The galaxy name and its luminosity in units of $L^*$ are given in each panel. 
In each case, the best-fit distribution obtained from Monte-Carlo simulations 
of hierarchical galaxy formation is shown by the dashed curve (C\^ot\'e $et~al.$ 2002). 
(Right Panel) Individual and combined mass spectra of proto-galactic fragments 
for the same 28 galaxies (dotted curves and open squares, respectively). The 
dashed line shows a Press-Schechter (1974) mass function with 
$M^* = 5\times10^{11}M_{\odot}$ and $n = -2$, where $n$ is the 
index of the cosmological power spectrum.}
\end{figure}

It was suggested by Harris \& Pudritz (1994) that GCs form within the
dense cores of supergiant molecular clouds (SGMCs): $i.e.$, an hypothesized 
population of pressure-confined, self-gravitating, isothermal, magnetized
clouds. These SGMCs, they argued, were supported against gravitational
collapse by magnetic field pressure and Alfv\'enic turbulence. Though most SGMCs
would presumably have been disrupted during the assembly of the host galaxy, 
any surviving SGMCs would, if left in isolation, 
have masses of $M \sim~10^9~M_{\odot}$ 
and diameters of $D \sim~1$~kpc. In short, they would resemble dwarf galaxies.

Subsequent cosmological N-body/TREESPH simulations by Weil \& Pudritz (2001) 
confirmed that gravitationally-bound objects having masses and sizes similar
to these SGMCs do indeed form,
and that they have a roughly power-law mass spectrum, $N(M) \propto M^{\alpha}dM$,
with $\alpha \sim -1.7$. Since star and GC formation was not included
in their simulations, no conclusion could be drawn regarding GC MDFs.

An empirical method of studying GC MDFs has been has been explored in a 
series of papers by my collaborators and I (C\^ot\'e $et~al.$ 1998, 2000, 2002) 
where a Monte-Carlo algorithm was developed to simulate the GC MDFs of 
galaxies that are assembled from ``proto-galactic fragments". In this picture, 
both the total number of GCs and their metallicities are determined solely by 
the mass of the proto-galactic fragment in which they formed; no GCs are 
formed during the merger and accretion process. The left panels of Figure~1 
show simulated and observed MDFs for nine of the 28 galaxies examined in 
C\^ot\'e $et~al.$ (2002). The right panel shows proto-galactic mass spectra 
for the 28 galaxies, along with the ensemble mass spectrum. These 
simulations suggest that the mass spectrum of proto-galactic fragments had 
an approximate power-law form with
index $\alpha \sim -1.8$, in close agreement with the N-body results.

It is remarkable that two completely independent lines of evidence ($i.e.$,
N-body simulations of the gravitational collapse of primordial density 
fluctuations and Monte-Carlo simulations of the GC MDFs)
point to a steep proto-galactic mass spectrum. 
Although this agreement lends credibility to the results, it is important
to recall that the inferred mass spectrum is {\it much} steeper than 
measured luminosity function of galaxies in the local universe. Thus,
the formation of GC systems seems inevitably linked to the so-called 
``missing satellite" problem (Klypin $et~al.$ 1999; Moore $et~al.$ 1999).

In a recent paper that further demonstrates the utility of GC systems
in constraining cosmological models of galaxy formation, Beasley 
$et~al.$ (2002) used the GALFORM semi-analytic code to simulate the GC 
systems of 450 elliptical galaxies. The relative number of MP and MR GCs 
in their simulations --- which 
include shock heating of the gas, radiative cooling, feedback, mergers 
and stellar evolution --- were then compared to the observed numbers in 
M49, where the census of GCs is very nearly complete. This comparison revealed
that two {\it ad hoc} assumptions were necessary to reproduce the MP and MR
GCs in the correct numbers. First, the formation efficiency of MR GCs
must exceed that of MP GCs by a factor of $\approx$ 3-4 (see below for
an examination of the plausibility of this assumption). Second, 
the formation of MP GCs, which occur in low-mass halos that Beasley $et~al.$ 
term ``proto-galactic
disks", must be truncated abruptly at $z_{\rm trunc} \simeq 5$. Without these
assumptions, the predicted population of MP GCs would dominate the
MR population by up to two orders of magnitude at $z = 0$. Unlike previous
simulations, Beasley $et~al.$ were able to estimate {\it ages} for the GCs.
The MP GCs have, by virtue of the assumed value of $z_{\rm trunc}$, a mean 
age of $T \simeq$ 12 Gyr. The MR GCs are then found to be 1--7 Gyr 
younger, with a mean age difference of 3 Gyr, although this too 
hinges on the assumed $z_{\rm trunc}$. 

\subsubsection{\bf Post-Galactic:} Building upon a suggestion by Schweizer (1987),
Ashman \& Zepf (1992) examined the possibility that a factor-of-two difference
in specific frequency, $S_N$, between normal elliptical and spiral galaxies 
($e.g.$, Harris 1991) might be explained if: (1) normal elliptical galaxies formed
as a result of major mergers of disk galaxies, and (2) the merger process led 
to the formation of new, presumably MR, GCs. There seems to be little doubt that 
a process of this sort has played in role in the formation of some elliptical
galaxies, in light of the detection of young clusters forming 
in interacting galaxies ($e.g.$, Whitmore $et~al.$ 2001). But is it plausible
that major mergers of disk-galaxy pairs is the {\it generic} mechanism for
the formation of normal elliptical galaxies, more than half of which 
show bimodal GC MDFs ($e.g.$, Larsen $et~al.$ 2001)?

The most comprehensive study of GC formation in disk mergers is that of
Bekki $et~al.$ (2002), who used an N-body/TREESPH code to investigate mergers
of pairs of gas-rich disk galaxies. These disks were {\it posited} to contain 
a pre-existing population of exclusively MP GCs. Using a simple 
pressure criterion for the onset of GC formation, Bekki $et~al.$ showed that 
the end-products of such mergers do indeed resemble elliptical galaxies, with 
distinct MP and MR GC subpopulations. The newly-formed MR GCs show a greater degree 
of concentration toward the center of the merger remnant, as is observed. 
Bekki $et~al.$ also examined the kinematic properties of the MP and MR GC
subpopulations, finding the MP component to show a large velocity dispersion
and a net rotation; by contrast, the MR component was found to show a lower 
dispersion and modest rotation. By examining the GC MDFs of the merged galaxy,
Bekki $et~al.$ also noted that the mean metallicity of the MR GCs far
exceeded those measured for MR GCs in real ellipticals.
As no plausible change in the assumed range of gas
metallicities, abundance gradients in the progenitor disks, or GC
formation criteria was able to alter this basic result, they
concluded that ``if most elliptical galaxies are formed by major mergers,
then it must have occurred at high redshift". 

\section{The Globular Cluster Systems of M87 and M49 Compared}

Let us now consider how the various models fare in explaining the GC systems of
real ellipticals. I will focus on a comparison 
of two thoroughly-studied ellipticals: M49 and 
M87, the first- and second-ranked members 
of the Virgo cluster, respectively. Although they differ in $V$-band luminosity by only 
$\sim$ 20\%, the GC specific frequency of M87 exceeds that of M49 by a factor of 2-3
(Harris 1991). Together, these galaxies form a ``second parameter" pair in terms
of their GC systems, and any viable formation scenario must account for the
observed differences and similarities.

\begin{figure}[t]
\plotfiddle{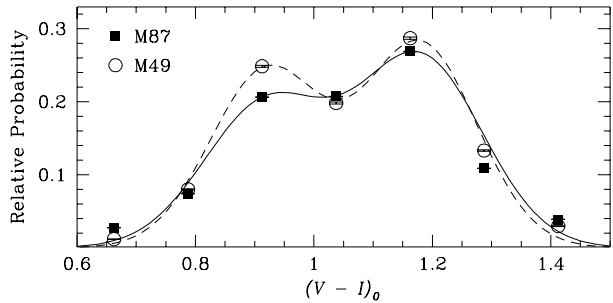}{1.5in}{0}{45}{45}{-150pt}{-80pt}
\caption{Color distributions for GCs belonging to M87 and M49, based on {\it HST} photometry
from Kundu $et~al.$ (1999) and Kundu \& Whitmore (2001). The solid and dashed curves
for the best-fit double Gaussians used to estimate the GC formation efficiencies
given in Table~1. The relative numbers of MP and MR GCs in the two galaxies is
remarkably similar, despite the factor of 2-3 difference in specific frequency.}
\end{figure}

The primary motivation of the Ashman \& Zepf (1992) merger model
was an explanation of the higher specific frequencies of early-type galaxies
relative to disk galaxies. In this picture, mergers serve to increase 
specific frequency through the formation of MR GCs. Thus, galaxies
with high specific frequencies --- of which M87 is the prototype --- are 
expected to have predominantly MR GC systems according to this model.
In the specific case of M87, this ``excess" population relative to M49
should amount to $\approx 7000$ {\it additional} MR GCs.

As Figure~3 shows, this expectation is not borne out by the observations. 
The measured GC color distributions for the two galaxies
(based on deep HST imaging in identical filters from Kundu $et~al.$
1999 and Kundu \& Whitmore 2001) are remarkably similar. Parameterizing each
distribution with a double-Gaussian yields the percentages of MP and
MR GCs given in Table~1. This table also give the total number of GCs belonging to
each galaxy, $N_{\rm tot}$, from McLaughlin (1999), who
showed that the observed difference in specific frequency between these 
two galaxies is likely due to the greater mass in X-ray emitting gas in the
vicinity of M87.

\begin{table}
{\centering
\caption{GC Formation Efficiencies for M87 and M49}
\medskip
\begin{tabular}{lcc}
\tableline
\tableline
 &  M87  &  M49  \\
\tableline
N$_{\rm tot}$        & 13,600$\pm$600    &     6,850$\pm$550 \\
$\epsilon_{\rm tot}$ & 0.28$\pm$0.04\% & 0.23$\pm$0.05\% \\
$f_{\rm MP}$         & 57$\pm$5\%     &     54$\pm$5\% \\
$f_{\rm MR}$         & 43$\pm$5\%     &     46$\pm$5\% \\
$\epsilon_{\rm MP}$  & 0.16$\pm$0.03\% & 0.12$\pm$0.03\% \\
$\epsilon_{\rm MR}$  & 0.12$\pm$0.02\% & 0.11$\pm$0.03\% \\
\tableline
\end{tabular}
\par}
\centering
\end{table}

Defining the GC formation
efficiency, $\epsilon_{\rm tot}$, as the total mass in GCs normalized to the total baryonic
mass ($i.e.,$ stars and gas), McLaughlin found the GC formation efficiencies
given in Table~1. The overall efficiency with which GCs formed in M87 and M49
are not only consistent with each other, they are indistinguishable from the ``universal"
GC formation efficiency of $\langle \epsilon_{\rm tot} \rangle  = 0.26\pm0.05$\% found by McLaughlin (1999)
from an analysis of nearly one hundred early-type galaxies (see also Blakeslee $et~al.$ 1997).
In any event, it is clear from Table~1 and Figure~1 that the similar
{\it relative} number of MP and MR GCs in the two galaxies is certainly at odds
with the predictions of the merger model.

Combining the values of $f_{\rm MP}$ and $f_{\rm MR}$ from Table~1 with
McLaughlin's best estimates for $\epsilon_{\rm tot}$ in the two galaxies, 
I find the values of $\epsilon_{\rm MP}$ and $\epsilon_{\rm MR}$ given in Table~1.
To within the uncertainties, the MP and MR GC systems have identical
formation efficiencies. The observed values of $\epsilon_{\rm MP}$ are 
roughly consistent with the formation efficiency of MP GCs, $\epsilon_{\rm MP} = 0.2$\%,
used by Beasley $et~al.$ (2002) in their semi-analytic simulations. However,
their assumed value of $\epsilon_{\rm MR} = 0.7$\% exceeds the measured 
value by a factor of six. Strictly speaking, the GC formation efficiencies in Table~1
are measured relative to the {\it baryonic} mass, whereas those of Beasley $et~al.$
are measured with respect to the {\it total} mass ($i.e.$, including dark matter). However, 
inside $\sim$ 50 kpc, the $R$-band mass-to-light ratios of M49 and M87 are virtually 
identical, with $\Upsilon_R \sim 35$ (C\^ot\'e $et~al.$ 2001; 2003), so the comparison 
should be valid in this case.

\begin{table}
\caption{Kinematics of GC Subpopulations in M87 and M49}
\medskip
\begin{tabular}{lccccccc}
\tableline
\tableline
 & \multicolumn{3}{c}{M87} & & \multicolumn{3}{c}{M49} \\
 \cline{2-4} \cline{6-8} \\
 & $\langle {\sigma}_p\rangle$ & $\langle {\Omega}R\rangle$ & $\langle {\Omega}R/{\sigma}_p\rangle$ & & $\langle {\sigma}_p \rangle$ & $\langle {\Omega}R \rangle$ & $\langle {\Omega}R/{\sigma}_p \rangle$ \\
 & (km~s$^{-1}$) & (km~s$^{-1}$) &  && (km~s$^{-1}$) & (km~s$^{-1}$) & \\
\tableline
All & 383$^{+31}_{-7}$ & 171$^{+39}_{-30}$ & 0.45$\pm0.09$ & & 316$^{+27}_{-8}$ & 48$^{+52}_{-26}$ & 0.15$^{+0.15}_{-0.08}$ \\
MP  & 397$^{+36}_{-14}$ & 186$^{+58}_{-41}$ & 0.47$^{+0.13}_{-0.11}$ & & 342$^{+33}_{-18}$ & 93$^{+69}_{-37}$ & 0.27$^{+0.19}_{-0.11}$ \\
MR  & 365$^{+38}_{-18}$ & 155$^{+53}_{-37}$ & 0.43$^{+0.14}_{-0.12}$ & & 265$^{+34}_{-13}$ &-26$^{+64}_{-79}$ & 0.10$^{+0.27}_{-0.25}$ \\
\tableline
\end{tabular}
\end{table}

Now let us examine the kinematical properties of the two GC systems.
A dynamical analysis of the M87 GC system based on Washington photometry
and radial velocities for 278 confirmed GCs was presented recently in
C\^ot\'e $et~al.$ (2001) and a similar analysis for the M49
GC system, based on velocities for 263 GCs, is now underway
(C\^ot\'e $et~al.$ 2003). Figure~2 shows the radial variation
in kinematic parameters ($i.e.$, velocity dispersion, $\sigma_p$, projected
rotation amplitude, $\Omega{R}$, position angle of the rotation axis,
$\Theta_0$, and the ratio of rotation amplitude to velocity dispersion,
$\Omega{R}/\sigma_p$) for the MP and MR GCs in M87. Global
results for the two galaxies are presented in Table~2.

A detailed discussion of the results is beyond the scope of 
this review, but the main findings can be summarized as follows. In
M87, both the MP and MR GC systems show significant rotation, with
$\langle {\Omega}R/\sigma \rangle \approx 0.45$. The MR GCs rotate about the galaxy's
minor axis everywhere, as do the MP GCs beyond $R \sim 15$ kpc. Inside
this radius, however, the MP GCs appear to rotate about the {\it major}
axis. In the case of M49, the MP GCs show modest minor axis rotation everywhere, with 
$\langle {\Omega}R/\sigma_p \rangle \simeq 0.27$,
while the MR GCs show little or no rotation. (The formal best-fit
actually suggests that the MR GCs are counter-rotating). 

\begin{figure}[t]
\plotfiddle{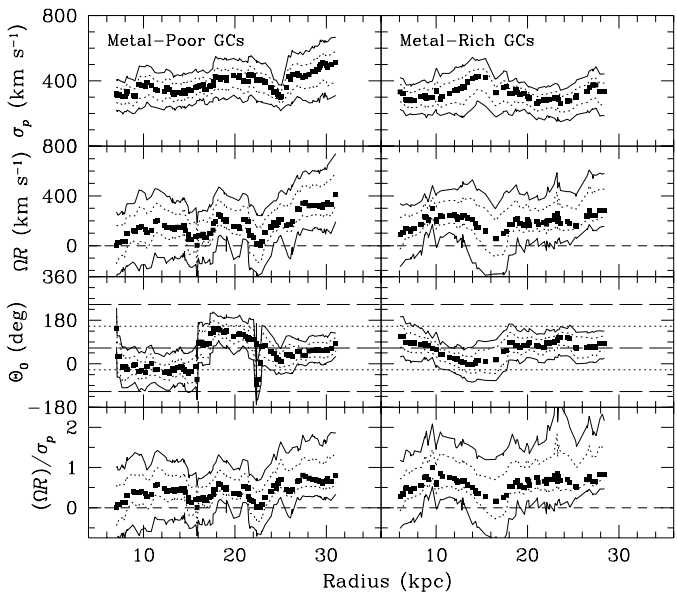}{3in}{0}{45}{45}{-150pt}{-80pt}
\caption{Kinematic properties of GCs in M87. The panels show smoothed profiles 
for the velocity dispersion, rotation amplitude, position angle of the rotation 
axis, and the ratio of rotation amplitude to line-of-sight velocity dispersion. 
The dashed and dotted lines in the third panels show the orientation of the
galaxy's minor and major axes, respectively.
Results for the MP GCs are shown on the left; MR GCs are on the right. Only 
those points that are separated by more than 90$^{\prime\prime}$ $\simeq$ 6.5~kpc 
are independent. The dotted and solid curves show the 68\% and 95\% confidence 
intervals on measured parameters.} 
\end{figure}

Clearly, the GC subsystems in M87 and M49 show some rather dramatic kinematic
differences. This apparent complexity becomes is all the more puzzling when one
considers that the two spiral galaxies have the best studied GC systems,
the Milky Way and M31, have $\langle {\Omega}R/\sigma \rangle \simeq 0.32$
and $\langle {\Omega}R/\sigma \rangle \simeq 0.85$ for the MP components,
respectively.  For their MR GC subpopulations, rotation is even more important,
with $\langle {\Omega}R/\sigma \rangle \simeq 1.05$ and
$\langle {\Omega}R/\sigma \rangle \simeq 1.10$ respectively
(C\^ot\'e 1999; Perrett $et~al.$ 2002).

What implications do these results have for GC formation models? First, the
rotation observed for MP GCs in these galaxies
suggests that, if these GCs did indeed originate in numerous, low-mass
proto-galactic fragments as suggested by several of the ``proto-galactic" models
discussed above, then these low-mass fragments must first have coalesced into
larger entities before being incorporated into
the final galaxy (assuming that the observed rotation reflects
the orbital angular momentum of the progenitor galaxies). Secondly,
the rapid rotation among the MR GCs in M87 (and in the two spirals) 
indicates that their formation in mergers may be unlikely, given that
angular momentum transport should produce a slowly rotating
population. On the other hand, the MR GCs in M49 are consistent with
no net rotation, suggesting that angular momentum transport during
a major merger may have been effective in this case.

Finally, I consider the critical issue of ages for the MP and MR
GCs subpopulations in these galaxies, focusing on two
complementary techniques that have been brought to bear upon
this issue: integrated-light spectroscopy, and the magnitude 
difference between the turnovers of the GC luminosity functions (LFs) for
the GC subpopulations. The former technique yields absolute ages
for the GC subpopulations, while the latter approach gives only
their relative ages.

The various age measurements are summarized in Table~3, which reports
$\Delta{T} \equiv T_{\rm MP} - T_{\rm MR}$ for each GC subpopulation and, 
when available, the corresponding absolute ages. In the case of M87,
the spectroscopy of Cohen $et~al.$ (1998) suggests an old age for 
both components, with no evidence for a trend between age and
metallicity. On the other hand, Kundu $et~al.$ (1999) found a rather 
large age difference from the $V$ and $I$ luminosity functions, with 
the MR GCs being several Gyr younger. A more recent application of
this technique, however, which relies on Str\"omgren photometry from
from {\it HST} and makes use of the greater age sensitivity of the $u$-band,
points to a very small age difference, in agreement with
the spectroscopic results (Jord\'an $et~al.$ 2002).
For M49, both techniques indicate that the GC subpopulations are 
coeval to within the measurement errors.

\begin{table}
\caption{Age Determinations for GC Subpopulations in M87 \& M49}
\medskip
\begin{tabular}{lccclc}
\tableline
\tableline
Galaxy & $T_{\rm MP}$ & $T_{\rm MR}$ & ${\Delta}T \equiv T_{\rm MP} - T_{\rm MR}$ & Source & Ref. \\
\tableline
M87 & 13.7$\pm$1.8 & 12.7$\pm$2.2 & +1.0$\pm$3.3    & Spectra & 1 \\
    & \nodata      & \nodata      & $\approx$3--6   & LF ($VI$) & 2 \\
    & \nodata      & \nodata      & +0.2$\pm$1.5(sys.)$\pm$2.0(ran.) & LF ($uvby$) & 3 \\
& & & & & \\
M49 & \nodata      & \nodata      & $-$0.6$\pm$3.2  & LF ($VI$) & 4 \\
    & 14.5$\pm$4   & 13.8$\pm$6   & +0.7$\pm$7.2    & Spectra & 5 \\
\tableline
\end{tabular}
\smallskip
REFERENCES: (1) Cohen $et~al.$ (1998);
(2) Kundu $et~al.$ (1999);
(3) Jord\'an $et~al.$ (2002);
(4) Puzia $et~al.$ (1999);
(5) Beasley $et~al.$ (2000).
\end{table}

The agreement between the photometric and spectroscopic results is 
generally encouraging. For these two galaxies at least, the MP and MR GCs
seem to be truly old objects. Although the error bars remain large,
it seems that, if GC formation in mergers of any sort --- minor or 
major --- has played the dominant role in producing the MR GCs in 
these galaxies, then the merger/assembly process must have occurred 
at high redshift ($i.e.$, $z \gae 4$ in the currently fashionable 
$\Lambda$CDM models). 
In my opinion, these results suggest that the disparate metallicities
of the GC subpopulations in these galaxies likely reflect {\it environmental}
differences in the local sites of GC formation.

\section{Conclusions}

In both M87 and M49, the efficiency of GC formation is found to have the 
familiar ``universal" value of $\epsilon_{\rm tot}~\simeq~0.25$\% 
per unit baryon mass (McLaughlin 1999). In both galaxies, the 
formation efficiencies of the separate MP and MR components is
roughly half that of the GC system as a whole. These findings 
provide a challenge to models in which MR GCs form exclusively in 
major mergers, and to models that require different 
formation efficiencies for the MP and MR GCs.

Kinematic studies of the GC subpopulations in M87 and M49 reveal some
interesting differences. In M87, both GC subpopulations show rapid
rotation; in M49, the MP GCs show modest rotation, while the MR GCs
form a non-rotating population. The existence of rapidly rotating GC 
subpopulations in these galaxies, and in the Milky Way and M31,
suggests that mergers have probably played a role in the formation 
of the host galaxies, as the amount of angular momentum involved
seems to exceed that generated by tidal torques alone (see C\^ot\'e
$et~al.$ 2001). With the exception of M49, the rapidly rotating
MR GCs in these galaxies indicate that the formation of the MR GCs 
probably predated the mergers. 
Furthermore, the ages of the GC subpopulations in M87
and M49 suggest that if the mergers were predominantly dissipational
in nature, then the mergers must have occurred at 
high redshift, as noted by Bekki $et~al.$ (2002) from an analysis
of simulated and observed GC MDFs.

Many properties of the MP GCs in M87 and M49 (and in other galaxies)
are consistent with the predictions of models in which
these GCs form in low-mass, proto-galactic fragments that are later 
accreted into the main body of the galaxy. 
These proto-galactic fragments 
appear to be the analogs of the low-mass subhalos that form in 
cosmological simulations of structure formation. Indeed, the literature
abounds with terminology for these objects: $e.g.$,
proto-galactic fragments, supergiant molecular clouds, 
proto-galactic disks, Searle-Zinn fragments, failed dwarfs, $etc$.
In terms of their physical properties, however, these are 
the same objects.

While there may be an emerging consensus on the origin of the MP GCs, debate
continues on the nature and origin of their MR counterparts. These clusters
can be interpreted simply as those that formed in the most massive proto-galactic 
fragments, or as the endproducts of a star formation burst triggered by 
gas-rich mergers. The former interpretation has the {\ae}sthetic advantage 
of requiring a single mechanism to explain both GC populations, while
the second scenario has strong support from observations of young 
GCs forming in local mergers. It is likely
that both processes have played a part in shaping the GC systems that
we observe today, and our task for the decade ahead is to determine
which, if either, of these processes has played a dominant role.

I thank Judy Cohen, John Blakeslee, Andres Jord\'an, Ron Marzke, 
Dean McLaughlin and Mike West for allowing me to present preliminary 
results from our ongoing collaborations.


\begin{references}

\reference Ashman, K.M., \& Zepf, S.E. 1992, ApJ, 384, 50

\reference Beasley, M.A., Baugh, C.M., Forbes, D.A., Sharples, R.M., \& Frenk, C.S. 2002, MNRAS, 333, 383

\reference Beasley, M.A., Sharples, R.M.;,Bridges, T.J.;,Hanes, D.A., Zepf, S.E., Ashman, K.M., \& 
Geisler, D. 2000, MNRAS, 318, 1249

\reference Bekki, K., Forbes, D.A., Beasley, M.A., \& Couch, W.J. 2002, astro-ph/0206008

\reference Blakeslee, J.P., Tonry, J.L., \& Metzger, M.R. 1997, AJ, 114, 482

\reference Cen, R. 2001, ApJ, 560, 592

\reference Cohen, J.G., Blakeslee, J.P., \& Ryzhov, A. 1998, ApJ 496, 808

\reference C\^ot\'e, P. 1999, AJ, 118, 406

\reference C\^ot\'e, P., Marzke, R.O., \& West, M.J. 1998, ApJ, 501, 554

\reference C\^ot\'e, P., Marzke, R.O., West, M.J., Minniti, D. 2000, ApJ, 533, 869

\reference C\^ot\'e, P., West, M.J., \& Marzke, R.O. 2002, ApJ, 567, 853

\reference C\^ot\'e, P., McLaughlin, D.E., Hanes, D.A., Bridges, T.J., Geisler, D.; Merritt, D., 
Hesser, J.E., Harris, G.L.H., \& Lee, M.G. 2001, ApJ, 559, 828

\reference Harris, W.E., \& Pudritz, R.E. 1994, ApJ, 420, 177

\reference Jord\'an, A., C\^ot\'e, P., West, M.J., \& Marzke, R.O. 2002, ApJ, in press.

\reference Klypin, A., Kravtsov, A.V., Valenzuela, O., \& Prada, F. 1999, ApJ, 522, 82

\reference Kundu, A., Whitmore, B.C., Sparks, W.B., Macchetto, F.D.,  Zepf, S.E., \& Ashman, K.M. 1999, ApJ, 513, 733

\reference Kundu, A., \& Whitmore, B.C. 2001, AJ, 121, 2950

\reference Moore, B., Ghigna, S., Governato, F., Lake, G., Quinn, T., Stadel, J., \& Tozzi, P. 1999, ApJ, 524, L19

\reference McLaughlin, D.E. 1999, AJ, 117, 2398

\reference Peebls, P.J.E., \& Dicke, R.H. 1968, ApJ, 154, 891

\reference Peebles, P.J.E. 1984, ApJ, 277, 470

\reference Perrett, K.M., Bridges, T.J., Hanes, D.A., Irwin, M.J., Brodie, J.P., Carter, D., 
Huchra, J.P., \& Watson, F.G. 2002, AJ, 123, 2490 

\reference Press, W.H., \& Schechter, P. 1974, ApJ, 187, 425

\reference Puzia, T.H., Kissler-Patig, M., Brodie, J. P., \& Huchra, J.P. 1999, AJ, 118, 2734

\reference Schweizer, F. 1987 in Nearly Normal Galaxies, ed. S.M. Faber (Springer, New York), p. 18

\reference Weil, M.L., \& Pudritz, R.E. 2001, ApJ, 556, 164

\reference Whitmore, B.C., Zhang, Q., Leitherer, C., Fall, S.M., Schweizer, F., \& Miller, B.W. 1999, AJ, 118, 1551

\end{references}
\end{document}